\documentclass{webofc}
\usepackage[varg]{txfonts}

\begin{document}

\title{Equilibration of a strongly interacting plasma: holographic \\analysis of local and nonlocal probes}

\author{\firstname{Loredana} \lastname{Bellantuono}\inst{1,2}\fnsep\thanks{\email{loredana.bellantuono@ba.infn.it}}
}

\institute{Dipartimento di Fisica, Universit\`a  di Bari,via Orabona 4, I-70126 Bari, Italy 
\and
           INFN, Sezione di Bari, via Orabona 4, I-70126 Bari, Italy 
          }

\abstract{
The relaxation of a strongly coupled plasma towards the hydrodynamic regime is studied by analyzing the evolution of local and nonlocal observables in the holographic approach. The system is driven in an initial anisotropic and far-from equilibrium state through an impulsive time-dependent deformation (quench) of the boundary spacetime geometry. Effective temperature and entropy density are related to the position and area of a black hole horizon, which has formed as a consequence of the distortion. The behavior of stress-energy tensor, equal-time correlation functions and Wilson loops of different shapes is examined, and a hierarchy among their thermalization times emerges: probes involving shorter length scales thermalize faster.
}

\maketitle

\section{Boundary sourcing}
\label{Boundary sourcing}
The deconfined and strongly-coupled plasma produced in relativistic heavy ion collisions experiments is characterized by an initial configuration which is dense, hot and highly anisotropic. Such system expands and cools down, reaching a hydrodinamic regime $\sim 1$ fm/c after the collision. A possible framework for studying the equilibration process is holographic QCD. This approach is based on Maldacena's conjecture, according to which a strongly coupled Super-Yang Mills gauge theory on a $4$-dimensional Minkowski space (boundary) is dual to a classical gravity theory in the space $AdS_{5}\times\mathcal{S}^{5}$, with $AdS_{5}$ the $5$-dimensional anti-de Sitter manifold (bulk) and $\mathcal{S}^{5}$ a $5$-dimensional sphere \cite{Maldacena}.
In order to mimic the effects of heavy ion collisions and reproduce the initial far-from-equilibrium state, a time-dependent deformation (quench) is introduced to the metric on the boundary geometry (boundary sourcing) \cite{Chesler}. When the quench becomes static, the system starts to evolve towards the hydrodynamic regime. The duality correspondence is employed to transfer such a description in the bulk geometry. The evolution of the system is obtained by solving the Einstein equations in the curved $5$-dimensional spacetime.
Let us set the $4$-dimensional coordinates $x^{\mu}=\left(x^{0},x^{1},x^{2},x^{3}\right)$, with $x^{3}=x_{\parallel}$ the direction in which ion collisions and plasma expansion occurs. Boost-invariance along that axis is imposed, as well as translational and rotational $O(2)$ invariance in the transverse plane $\textbf{x}_{\perp}=\left(x_{1},x_{2}\right)$. The above symmetries suggest us to introduce the proper time $\tau$ and the spacetime rapidity $y$, defined through the relations $x^{0}=\tau \mathrm{cosh} y$ and $x_{\parallel}=\tau \mathrm{sinh} y$. The $4$-dimensional Minkowski line element $ds_{4}^{2}=-d\tau^{2}+dx_{\perp}^{2}+\tau^{2}dy^{2}$ is distorted impulsively:
\begin{equation}\label{eq: 4d}
ds_{4}^{2}=-d\tau^{2}+e^{\gamma(\tau)}dx_{\perp}^{2}+\tau^{2}e^{-2\gamma(\tau)}dy^{2}\,;
\end{equation}
note that this deformation leaves the spatial three-volume invariant and respects the aforementioned symmetries. The $5$-dimensional dual metric can be parametrized in terms of Eddington-Finkelstein coordinates as
\begin{equation}\label{eq: 5d}
ds^2_5=2 dr d\tau-A d\tau^2+ \Sigma^2 e^B dx_\perp^2+ \Sigma^2 e^{-2B}dy^2  \,,
\end{equation}
with $r$ the fifth holographic coordinate. The boundary of this geometry is obtained in the limit $r\to\infty$ and corresponds to the $4$-dimensional spacetime \eqref{eq: 4d}. The metric functions $A$, $B$ and $\Sigma$ depend only on $r$ and $\tau$ because of the imposed symmetries. They are solutions of Einstein's equations with negative cosmological constant, that can be rephrased as \cite{Chesler}:
\begin{equation}\label{eq: einstein 1}
\Sigma ({\dot \Sigma})^\prime +2 \Sigma^\prime {\dot \Sigma}-2 \Sigma^2 = 0
\end{equation}
\begin{equation}\label{eq: einstein 2}
\Sigma ({\dot B})^\prime+\frac{3}{2} \left(\Sigma^\prime {\dot B}+B^\prime {\dot \Sigma}\right) = 0
\end{equation}
\begin{equation}\label{eq: einstein 3}
A^{\prime \prime} +3 B^\prime {\dot B} -12 \frac{\Sigma^\prime {\dot \Sigma} }{\Sigma^2}+4 = 0
\end{equation}
\begin{equation}\label{eq: einstein 4}
{\ddot \Sigma}+\frac{1}{2} \left( {\dot B} ^2 \Sigma -A^\prime {\dot \Sigma} \right) = 0
\end{equation}
\begin{equation}\label{eq: einstein 5}
\Sigma^{\prime \prime}+\frac{1}{2} B^{\prime 2} \Sigma = 0 \,.
\end{equation}
The notations $\xi^\prime\equiv\partial_r \xi$ and ${\dot \xi}\equiv \partial_\tau \xi+ (A/2) \partial_r \xi$ for a generic function $\xi(r,\tau)$ indicate directional derivatives along the infalling radial null geodesics and the outgoing radial null geodesics, respectively. The system \eqref{eq: einstein 1}--\eqref{eq: einstein 5} involves partial differential equations that should be solved recursively\,: an efficient procedure for the resolution is discussed in detail in \cite{Bellantuono 2015}. $A$, $B$ and $\Sigma$ are determined by imposing two boundary conditions:
\begin{itemize}
\item As $r\to\infty$ the metric \eqref{eq: 5d} reduces to \eqref{eq: 4d};
\item At the initial time $\tau=\tau_{i}$, when the quench is turned on, the bulk metric coincides with the $AdS_{5}$ form
\begin{equation}\label{eq: AdS5}
ds^2= r^2 \left[ -d \tau^2+ d x_\perp^2 + \left( \tau+\frac{1}{r} \right)^2 dy^2 \right] + 2 dr d\tau \, . 
\end{equation}
\end{itemize}
A strategy to monitor the plasma during thermalization is to compute characteristic probes in the boundary sourcing approach and to compare them with the corresponding viscous hydrodynamic expressions. A quench profile $\gamma(\tau)$ chosen for the analysis is shown in the top left panel of figure~\ref{fig-1}; it can be viewed as the overlap of two signals having different time scales: a short pulse ending at $\tau_{f}=5$ and a step function approaching a constant value.

\section{Viscous hydrodynamic regime}
\label{Viscous hydrodynamic regime}
Let us present an overview on viscous hydrodynamics in a homogeneous $4$-dimensional spacetime, whose metric is invariant under boosts along $x_{\parallel}=x_{3}$, and under translations and rotations in the $\textbf{x}_{\perp}=\left(x_{1},x_{2}\right)$ plane. We focus on the stress-energy tensor $T^{\mu}_{\nu}$, whose components are the energy density $\varepsilon$ and the pressures transverse and longitudinal with respect to the collisional axis, $p_{\perp}$ and $p_{\parallel}$.
As a consequence of the imposed symmetries, these quantities depend only on the proper time $\tau$ \cite{Bjorken:1982qr}. Moreover, a conserved and traceless stress-energy tensor can be expressed in terms of a single function $f(\tau)$ as 
\begin{equation}\label{eq: Tmunu hydro}
T_{\mu}^{\nu}(\tau)=diag \left(-f(\tau),\,f(\tau)+\frac{1}{2}\tau f^\prime(\tau),\,f(\tau)+\frac{1}{2}\tau f^\prime(\tau),\, -f(\tau)-\tau f^\prime(\tau)\right) \, .
\end{equation}
For a perfect fluid, the equation of state $\varepsilon=3 p$ and the isotropy condition $p=p_\perp=p_\parallel$ fix the $\tau$-dependence: $\varepsilon(\tau)\propto \tau^{-4/3}$, which is modified if viscous effects are included \cite{Janik:2005zt}. The effective temperature $T_{eff}(\tau)$, defined through the relation $\varepsilon(\tau)= 3\pi^4 T_{eff}(\tau)^{4}/4$, has been computed in a late-time expansion of $\mathcal{N}=4$ SYM \cite{Heller:2007qt}, with the result
\begin{eqnarray}\label{eq: Teff hydro}
T_{eff}(\tau)&=&\frac{\Lambda}{(\Lambda \tau)^{1/3}} \Bigg[ 1-\frac{1}{6 \pi (\Lambda \tau)^{2/3}}+\frac{-1+\log 2}{36 \pi^2 (\Lambda \tau)^{4/3} }
+\frac{-21+2\pi^2+51 \log 2 -24 (\log 2)^2}{1944 \pi^3 (\Lambda \tau)^2} \nonumber \\ 
&+& {\cal O}\left( \frac{1}{(\Lambda \tau)^{8/3}} \right )\Bigg] \, ,
\end{eqnarray}
The corresponding viscous hydrodynamic expressions for the components of $T^{\mu}_{\nu}$ at large $\tau$ are:
\begin{equation}\label{eq: epsilon hydro}
\varepsilon(\tau) = \frac{3 \pi^4 \Lambda^4}{4 (\Lambda \tau)^{4/3} }\left[ 1-\frac{2c_1}{ (\Lambda \tau)^{2/3}}+\frac{c_2}{ (\Lambda \tau)^{4/3}} + {\cal O}\left( \frac{1}{(\Lambda \tau)^2} \right )\right] \, , 
\end{equation}
\begin{equation}
p_\perp (\tau) = \frac{ \pi^4 \Lambda^4}{ 4(\Lambda \tau)^{4/3} } \left[ 1-\frac{c_2}{ (\Lambda \tau)^{4/3}} + {\cal O}\left( \frac{1}{(\Lambda \tau)^2} \right ) \right] \, ,\label{eq: pperp hydro}
\end{equation}
\begin{equation}\label{eq: ppar hydro}
p_\parallel (\tau) = \frac{ \pi^4 \Lambda^4}{ 4(\Lambda \tau)^{4/3} } \left[ 1-\frac{6c_1}{ (\Lambda \tau)^{2/3}}+\frac{5c_2}{ (\Lambda \tau)^{4/3}} + {\cal O}\left( \frac{1}{(\Lambda \tau)^2} \right )\right] \, ,
\end{equation}
with $c_1=1/(3 \pi)$ and $c_2=(1+2 \log{2})/(18 \pi^2)$. The parameter $\Lambda$ appearing in Eqs. \eqref{eq: Teff hydro}--\eqref{eq: ppar hydro} will be fixed in Sect.~\ref{Thermalization probes}. A bulk metric reproducing via holographic renormalization the viscous hydrodynamic stress-energy tensor has been identified \cite{vanderSchee:2012qj,Bellantuono 2016}.

\section{Thermalization probes}
\label{Thermalization probes}
\begin{figure}[h]
\centering
\vskip-1cm
\includegraphics[width=0.8\textwidth]{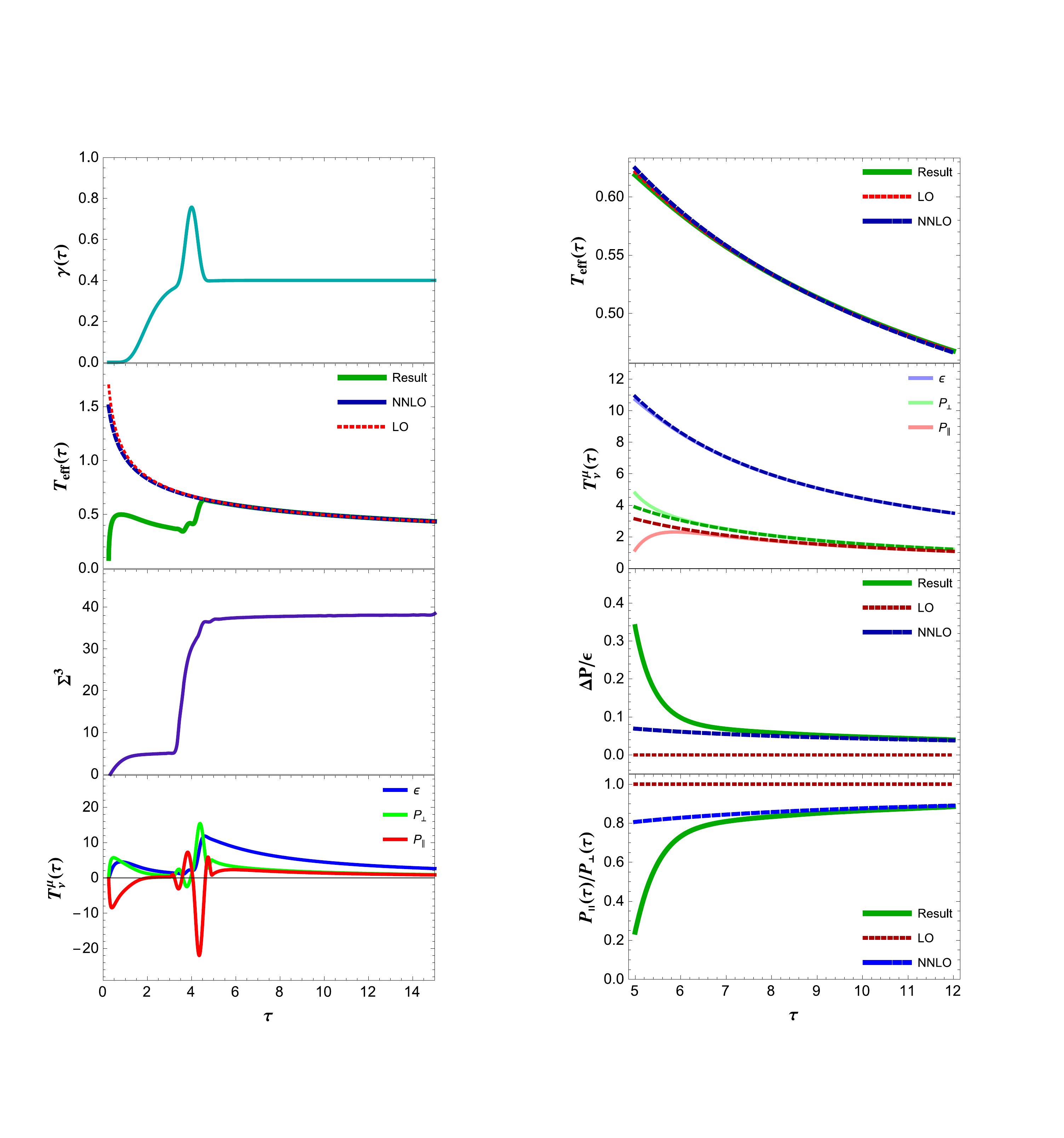}
\vskip-1cm
\caption{Left, from top to bottom: (i) quench profile $\gamma(\tau)$, (ii) temperature $T_{eff}(\tau)$, (iii) horizon area per unit rapidity $\Sigma^{3}\left(r_{h}(\tau),\tau\right)$ and (iv) the three components $\varepsilon(\tau)$, $p_{\perp}(\tau)$ and $p_{\parallel}(\tau)$ of the stress-energy tensor. Right, from top to bottom: (i) $T_{eff}(\tau)$, (ii) $\varepsilon(\tau)$, $p_{\perp}(\tau)$ and $p_{\parallel}(\tau)$, (iii) the pressure anisotropy $\Delta p/\varepsilon=\left(p_{\perp}-p_{\parallel}\right)/\varepsilon$ and (iv) the ratio $p_{\parallel}/p_{\perp}$ as compared to the corresponding hydrodynamic forms (LO and NNLO results in the $1/\tau$ expansion), at times after the end of the quench pulse \cite{Bellantuono 2015}.}
\label{fig-1}       
\end{figure}

As it can be deduced by solving Einstein's equations, the boundary distorsion $\gamma(\tau)$ determines a time-dependent horizon $r=r_{h}(\tau)$ in the bulk geometry \eqref{eq: 5d}. The gravitational radiation produced by the quench propagates in the fifth dimension and induces the creation of a black hole, whose horizon is $r_{h}(\tau)$ \cite{Chesler}. 
We can thus define, following the prescriptions of the holographic principle, the effective temperature and entropy of the system, which are proportional to the horizon position and area, respectively. Note that the quantity $\Sigma^{3}\left(r_{h},\tau\right)$ coincides with the horizon area per unit rapidity, and thus it can be used to characterize the entropy density. As shown in the left column of figure~\ref{fig-1}, $T_{eff}$ and $\Sigma^{3}$ manifest, respectively, a decreasing and a constant behavior in the time interval before the impulsive part of the quench $\gamma$. 
The stress-energy tensor on the boundary, $T^{\mu}_{\nu}(\tau)$, has been computed by the holographic renormalization procedure from the explicit metric functions
$A(r,\tau)$, $B(r,\tau)$ and $\Sigma(r,\tau)$ in \eqref{eq: 5d}. As a result of comparing its components $\varepsilon(\tau)$, $p_{\perp}(\tau)$ and $p_{\parallel}(\tau)$ with the corresponding hydrodynamic expressions \eqref{eq: epsilon hydro}--\eqref{eq: ppar hydro}, the parameter $\Lambda=1.12$ has been fitted. The results for such observables are displayed in figure~\ref{fig-1}.

\begin{figure}[h]
\centering
\vskip-2cm
\includegraphics[width=0.8\textwidth]{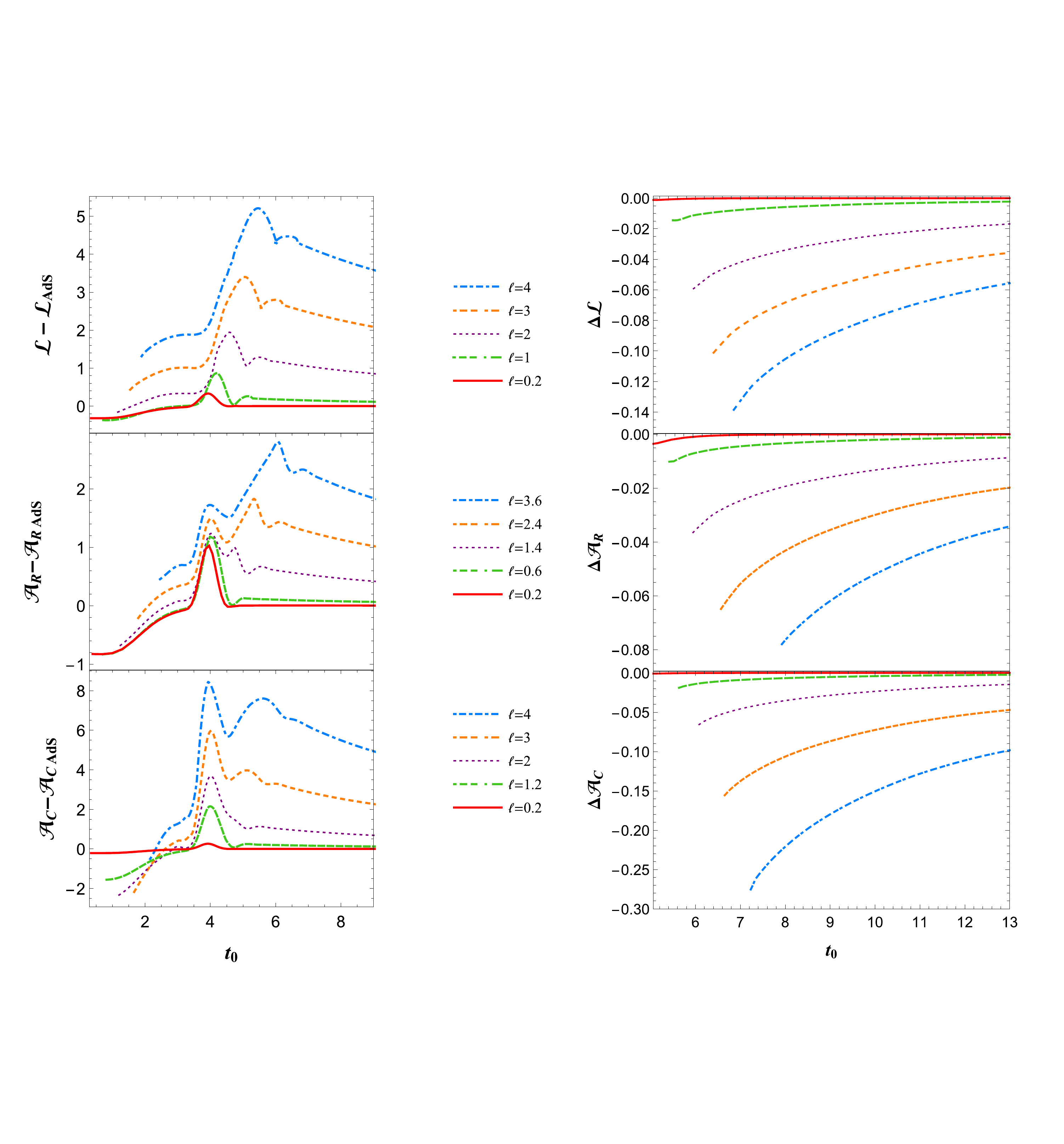}
\vskip-2cm
\caption{Left, from top to bottom: regularized lengths of the geodesics, regularized areas of the minimal surfaces for rectangular and circular Wilson loops. Regularization has been obtained subtracting from $\mathcal{L},\mathcal{A}_{R},\mathcal{A}_{C}$ the corresponding $AdS_{5}$ expressions computed from Eq. \eqref{eq: AdS5}.
Right, from top to bottom: differences between the observables $\mathcal{L},\mathcal{A}_{R},\mathcal{A}_{C}$ in the quenched geometry and the ones obtained using the hydrodynamic metric \cite{Bellantuono 2016}.}
\label{fig-2}
\end{figure}

Nonlocal observables allowing to examine the evolution of a strongly-coupled plasma are the two-point correlation function of boundary theory operators, and expectation values of Wilson loops defined on the boundary~\cite{Balasubramanian}. Let us consider a boundary scalar operator $\mathcal{O}$ having conformal dimension $\Delta\gg 1$. In AdS/CFT, its correlation function between two equal-time points with a spatial separation $\ell$ along one of the transverse directions, $P=\left(t_{0},-\ell/2,x_{2},y\right)$ and $Q=\left(t_{0},\ell/2,x_{2},y\right)$, can be written as 
\begin{equation}\label{eq: geo approximation}
\langle \mathcal{O} (t,\textbf{x}) \mathcal{O} (t,\textbf{x}') \rangle \simeq \sum_{\mathrm{geodesics}} e^{-\mathcal{L}\,\Delta} \, .
\end{equation}
$\mathcal{L}$ is the geodesic length, i.e. the length of the parametrized extremal trajectory $x^{M}(\lambda)$ ($x_{2}$ and $y$ fixed) connecting $P$ and $Q$:
\begin{equation}\label{eq: geo length}
\mathcal{L} = \int_{P}^Q d\lambda\sqrt{\pm g_{MN}\bar{x}^{M} \bar{x}^{N}} \, ,
\end{equation}
with $\bar{x}^{M}\equiv dx^{M}/d\lambda$, $g_{MN}$ the bulk metric, and $\pm$ signs in the square root for a space-like or time-like curve, respectively.
The approximation \eqref{eq: geo approximation} is reliable only for boundary operators with $\Delta\gg 1$. The top panel of figure~\ref{fig-2} displays the regularized length $\mathcal{L}-\mathcal{L}_{AdS}$, with $\mathcal{L}_{AdS}$ the geodesics length in the $AdS_{5}$ geometry \eqref{eq: AdS5}, as a function of the boundary time $t_{0}$. This quantity follows the quench profile $\gamma\left(t_{0}\right)$ with a delay that increases with the distance $\ell$ between the two points in the correlation function. A nonlocal probe of thermalization is the difference $\Delta\mathcal{L}$ between the geodesics length in the distorted geometry and the one computed in a bulk metric reproducing the viscous hydrodynamic behavior \eqref{eq: epsilon hydro}--\eqref{eq: ppar hydro} of $T^{\mu}_{\nu}$ \cite{vanderSchee:2012qj,Bellantuono 2016}. The top right panel of figure~\ref{fig-2} shows the relaxation of $\Delta\mathcal{L}$ for several sizes $\ell$, starting from the value of the physical time $\tilde{t}_{0}(\ell)$ at which it is no longer affected by the quench.
The expectation value of the Wilson loop along a closed path $\mathcal{C}$ of linear length scale $\ell$, fixed time $t_{0}$ and living on the boundary, can be computed through a similar approximation:
\begin{equation}
\langle W_{\mathcal{C}} \rangle \simeq e^{-\mathcal{A}\left(t_{0},\ell\right)},\quad\text{with}\quad\mathcal{A}\left(t_{0},\ell\right)=\frac{1}{2\pi\alpha'}\int d^{2}\xi \sqrt{det\left[g_{MN}\partial_{\alpha}x^{M}\partial_{\beta}x^{N}\right]} \, .
\end{equation}
$\mathcal{A}\left(t_{0},\ell\right)$ is the Nambu-Goto action, representing the area of the extremal surface bounded by $\mathcal{C}$ on the boundary and plunging in the bulk at fixed $y$, while $\xi^{\alpha}$ $(\alpha,\beta=1,2)$ and $x^{M}\left(\xi^{\alpha}\right)$ are respectively the worldsheet coordinates and the embedding of the surface into the bulk. Let us consider two different contours $\mathcal{C}$ of linear size $\ell$: a rectangle of finite width $\ell$ along the $x_{1}$ axis and infinite length along $x_{2}$, and a circumference of diameter $\ell$ in the transverse plane $\textbf{x}_{\perp}=\left(x_{1},x_{2}\right)$. The regularized areas of the extremal surfaces bounded by these paths have been computed in the boundary sourced geometry; the hydrodynamization of such quantities can be monitored through the nonlocal probes $\Delta\mathcal{A}_{\mathrm{R}}$ and $\Delta\mathcal{A}_{\mathrm{C}}$. The results, resumed in figure~\ref{fig-2}, are similar to the ones discussed for the geodesics.

\section{Hierarchy among thermalization times}
\label{Hierarchy among thermalization times}
The right columns in figures~\ref{fig-1} and~\ref{fig-2} display the physical observables of the distorted geometry in comparison with their hydrodynamic forms. The results can be summarized as follows.

The effective temperature $T_{eff}(\tau)$ reaches the hydrodynamic regime as soon as the boundary deformation becomes stationary.
Since the energy density is related to the effective temperature, it also acquires the hydrodynamical form at $\tau_f=5$, when the deformation pulse is turned off and $\gamma(\tau)$ approaches a constant value. On the other hand, pressure isotropy is restored with a time delay $\tau_p-\tau_f=1.74$, where the isotropization time $\tau_{p}=6.74$ has been identified as the value at which the computed ratio $p_{\parallel}/p_{\perp}$ differs from its hydrodynamical analogue by less than 5$\%$. These results can be expressed in physical units introducing an energy scale in the system. In particular, if the effective temperature at the end of the impulsive quench is set to $T_{eff}=500$ MeV, the pressures reach a common value after a time interval of $0.42$ fm/c. Such a delay is comparable to the values inferred from phenomenological analyses in heavy ion collisions experiments.

\begin{figure}[h]
\centering
\includegraphics[width=0.4\textwidth]{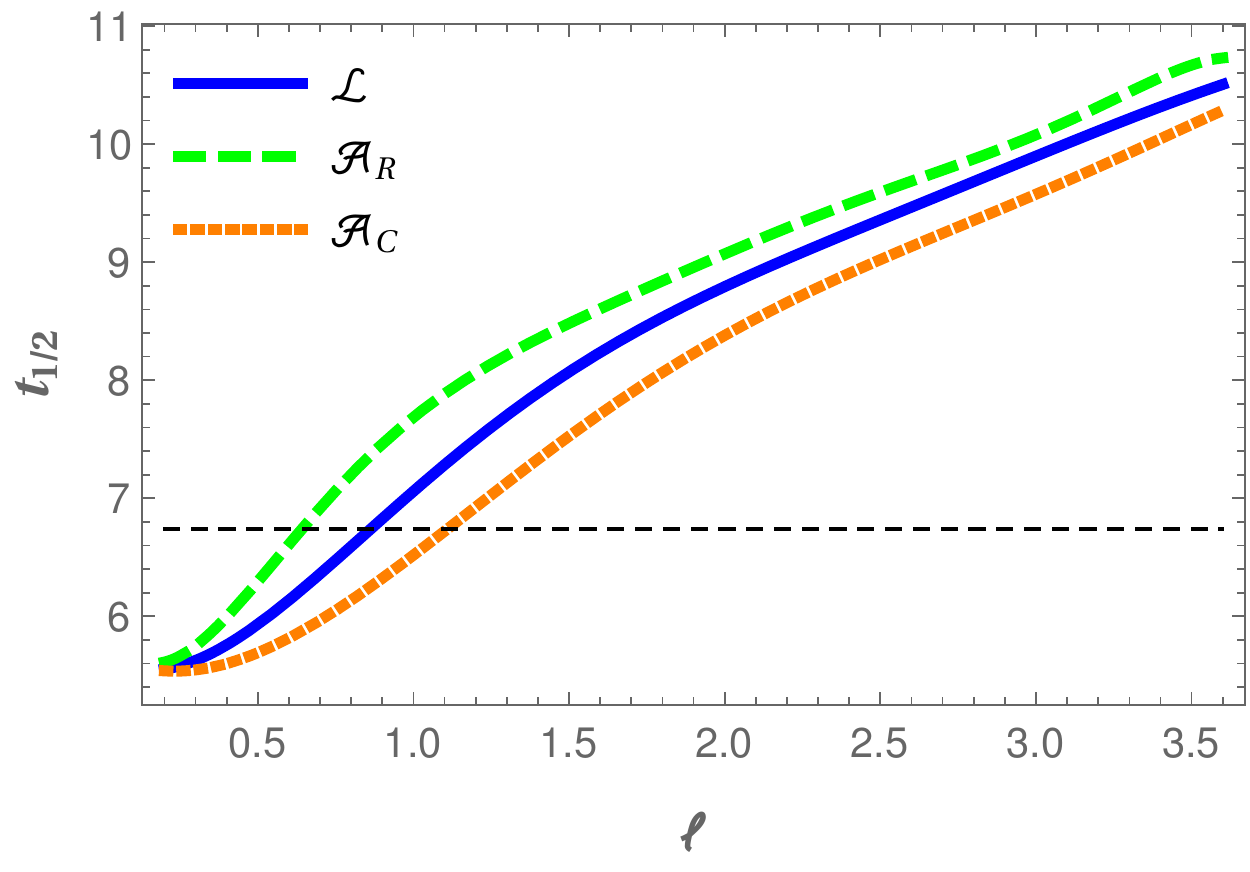}
\caption{Half-life $t_{1/2}(\ell)$ as a function of the size $\ell$ for the three nonlocal probes. The horizontal dashed line indicates the pressure isotropization time $\tau_{p}=6.74$ \cite{Bellantuono 2016}.}
\label{fig-3}
\end{figure}

The time a nonlocal probe needs for reaching the viscous hydrodynamic behavior depends on its size: for larger boundary separation between the geodesics extremes or among the contour points, the geodesics length and the areas bounded by Wilson loops take longer to equilibrate. It is interesting to evaluate the half-life $t_{1/2}(\ell)$, defined as the value of $t_{0}$ at which the $\Delta$-probes are reduced by a half with respect to their values at time $\tilde{t}_{0}(\ell)$, when the effect of the quench pulse on them vanishes. As shown in figure~\ref{fig-3}, the half-lives of the considered nonlocal probes exceed the pressure isotropization time $\tau_{p}$ for boundary separations $\ell\simeq1$, and increase linearly with $\ell$ for larger sizes. The hierarchy among the hydrodinamization times of the energy density, pressures and large probes has been identified, and shows that thermalization proceeds top-down, from short to large distances: UV thermalizes first. Such analysis can be extended to further nonlocal observables, e.g. the entanglement entropy \cite{Ecker:2015kna,Ecker:2016thn}.

{\bf \noindent Acknowledgments.}\\
I thank P. Colangelo, F. De Fazio, F. Giannuzzi and S. Nicotri for collaboration.

\end{document}